\documentclass[12pt]{iopart}

\usepackage{graphicx}
\usepackage{here}

\usepackage{stackrel}
\usepackage{amssymb}

\newcommand{\SU}{{\textrm{SU}}}

\newcommand{\vx}{{\mathbf{x}}}
\newcommand{\vr}{{\mathbf{r}}}
\newcommand{\vp}{{\mathbf{p}}}

\newcommand{\MeV}{\,{\rm MeV}}
\newcommand{\GeV}{\,{\rm GeV}}
\newcommand{\fm}{\,{\rm fm}}
\newcommand{\QCD}{{\textrm{\tiny QCD}}}
\newcommand{\HRG}{{\textrm{\tiny HRG}}}

\newcommand{\cons}{{\textrm{\scriptsize cons}}}
\newcommand{\lat}{{\textrm{\scriptsize lat}}}
\newcommand{\WKB}{{\textrm{\tiny WKB}}}

\newcommand{\PP}{{\mathcal P}}
\newcommand{\ignore}[1]{}
\newcommand{\nns}{{\textrm{\scriptsize ns}}}

\begin{document}

\title[Quark-diquark string tension, excited baryonic resonances
  and thermal fluctuations]{Quark-diquark string tension, excited
  baryonic resonances and thermal fluctuations}

\author{E.~Meg\'{\i}as, E.~Ruiz Arriola and L.L.~Salcedo}

\address{Departamento de F\'{\i}sica At\'omica, Molecular y Nuclear and Instituto Carlos I de F\'{\i}sica Te\'orica y Computacional, Universidad de Granada, \\ 
Avenida de Fuente Nueva s/n, 18071 Granada, Spain}
\ead{emegias@ugr.es, earriola@ugr.es, salcedo@ugr.es}
\vspace{10pt}

\begin{abstract}
We study the baryonic fluctuations from second to eighth order
involving electric charge, baryon number and strangeness below the
quark-gluon plasma crossover and numerically known from lattice QCD
calculations.  By considering a particular realization of the Hadron
Resonance Gas model, we provide evidence on the dominant role of
quark-diquark degrees of freedom to describe excited baryonic
resonances. After proving by means of suitable Polyakov loop
correlators that the quark-diquark and the quark-antiquark forces
coincide, $V_{\bar q q } (r) = V_{\bar q D } (r) + \textrm{const}$, we find that the
corresponding susceptibilities can be saturated with excited baryonic
states in a quark-diquark model picture.
\end{abstract}

%

\section{Introduction}
\label{sec:introduction}

The experimental program on relativistic heavy ion collisions is
currently ongoing at LHC within the ALICE
experiment~\cite{Citron:2018lsq}, and at RHIC in the so called
beam-energy scan (BES) program and further upgraded
stages~\cite{Odyniec:2019kfh}. In these experiments they study various
regions of the $T-\mu_B$ phase diagram by changing the collision
energy. There are also several future experiments at
FAIR~\cite{Senger:2017nvf} and NICA~\cite{Kekelidze:2017tgp} that will
contribute as well to the study of the Quantum Chromodynamics (QCD)
phase diagram with finite chemical potential.

Apart from lattice calculations~\cite{Borsanyi:2011sw,Bazavov:2012jq},
there are also further approaches to study QCD at finite temperature
and chemical potential, and in particular fluctuations of conserved
charges.  Among them, we could mention the chiral quark models coupled
to the Polyakov loop, as e.g. the Polyakov-loop Nambu--Jona-Lasinio
(PNJL) model~\cite{Shao:2017yzv}, and the Polyakov-loop quark meson
model~\cite{Wen:2018nkn}, as well as the Dyson-Schwinger
approaches~\cite{Isserstedt:2019pgx}. Another interesting approach is
the the Hadron Resonance Gas (HRG) model that allows to describe the
confined phase of the QCD equation of state (EoS) at finite
temperature below $\sim 150 \MeV$ as a multicomponent gas of
non-interacting massive stable and point-like
particles~\cite{Hagedorn:1965st,Hagedorn:1984hz}, which are usually
taken as the conventional hadrons listed in the review by the Particle
Data Group (PDG)~\cite{Tanabashi:2018oca}. A clear advantage of the
HRG approach is that it incorporates, by construction, the relevant
degrees of freedom that allow to describe the low temperature regime
of QCD. The other approaches cannot describe such regime so
accurately, and they usually focus on the description of the high
temperature and the phase transition regimes. These regions in the
phase diagram of QCD, however, cannot be described within the HRG
approach. In this sense, some authors have proposed some hybrid models
that incorporate the advantages of different approaches in a single
picture, as e.g. the HRG-PNJL hybrid
model~\cite{Miyahara:2019zfn}. Let us point out that all these works
try to describe lattice data.

One of the greatest achievements of the HRG approach has been the
study of the trace anomaly, $(\epsilon-3P)/T^4$ with $\epsilon$ energy
density and $P$ the pressure. It was computed directly in lattice QCD
by several collaborations~\cite{Borsanyi:2013bia,Bazavov:2014pvz} and
within the HRG approach, leading to an excellent agreement for
temperatures below $T \sim 170 \MeV$ (see
e.g. Ref.~\cite{Arriola:2014bfa} and references therein). Thus, this
approach has emerged as a practical and viable path to establish
completeness of hadronic states in the hadronic
phase~\cite{RuizArriola:2016qpb,Megias:2017qil,Megias:2016onb}. Apart
from the EoS, other thermal observables like the fluctuations of
conserved charges~\cite{Bazavov:2012jq} can be used to study the QCD
spectrum by distinguishing between different flavor sectors. Another
motivation to study the fluctuations is that they are among the most
relevant observables for finite density studies, as one possible way
to extend lattice results to finite density is to perform Taylor
expansions of the thermodynamical quantities around zero chemical
potential~\cite{Asakawa:2015ybt}. These quantities can directly be
compared to experimental measurements of fluctuations in relativistic
heavy ion collisions, and may provide insight into the existence of a
critical end point in the QCD phase
diagram~\cite{Ding:2015ona,Huovinen:2017ogf}.

There are a number of quark models that try to reproduce the PDG
hadron spectrum, one of the most fruitful being the Relativized Quark
Model (RQM) for mesons~\cite{Godfrey:1985xj} and
baryons~\cite{Capstick:1986bm}. This model shows that there are {\it
  further states} in the spectrum above some scale as compared to the
PDG. On the other hand, the suspicion that the baryonic spectrum can
be understood in terms of quark-diquark degrees of
freedom~\cite{Anselmino:1992vg} is rather old, and this includes
diquark clustering studies~\cite{Fleck:1988vm} as well as non
relativistic and relativistic
analyses~\cite{Santopinto:2004hw,Ferretti:2011zz,DeSanctis:2011zz,Galata:2012xt,Santopinto:2014opa,DeSanctis:2014ria,Gutierrez:2014qpa}. Lattice
QCD has also provided some evidence on diquarks correlations in the
nucleon~\cite{Alexandrou:2006cq}. In the present work we provide
further evidence for the existence of the quark-diquark baryonic
states, and study the possibility that these states saturate the
baryonic fluctuations in the confined phase as compared to the
available lattice QCD calculations.

  Finally, let us summarize the wider context of the research that we
  will pursue in this work. The phase transition from hadronic matter
  to a quark gluon plasma has been a major topic as it seems to be a
  distinct feature of QCD which may be produced copiously in
  ultrarelativistic heavy ion colliders and can be simulated by
  lattice QCD.  The thermal behaviour of the system is characterized
  by the equation of state which even in the conventional hadronic
  phase may provide valuable information on the hadronic
  spectrum. More specific information probing particular quantum
  numbers can be pinned down by analyzing fluctuations of conserved
  charges.  In this work we analyze the baryonic spectrum in terms of
  baryonic susceptibilities and find by comparison with lattice QCD
  calculations, PDG listings and quark models that most of the
  baryonic spectrum befits a quark-diquark picture.

\section{Hadron spectrum}
\label{sec:Hadron_spectrum}

\begin{figure*}[t]
\centering
 \begin{tabular}{c@{\hspace{1.5em}}c}
 \includegraphics[width=0.43\textwidth]{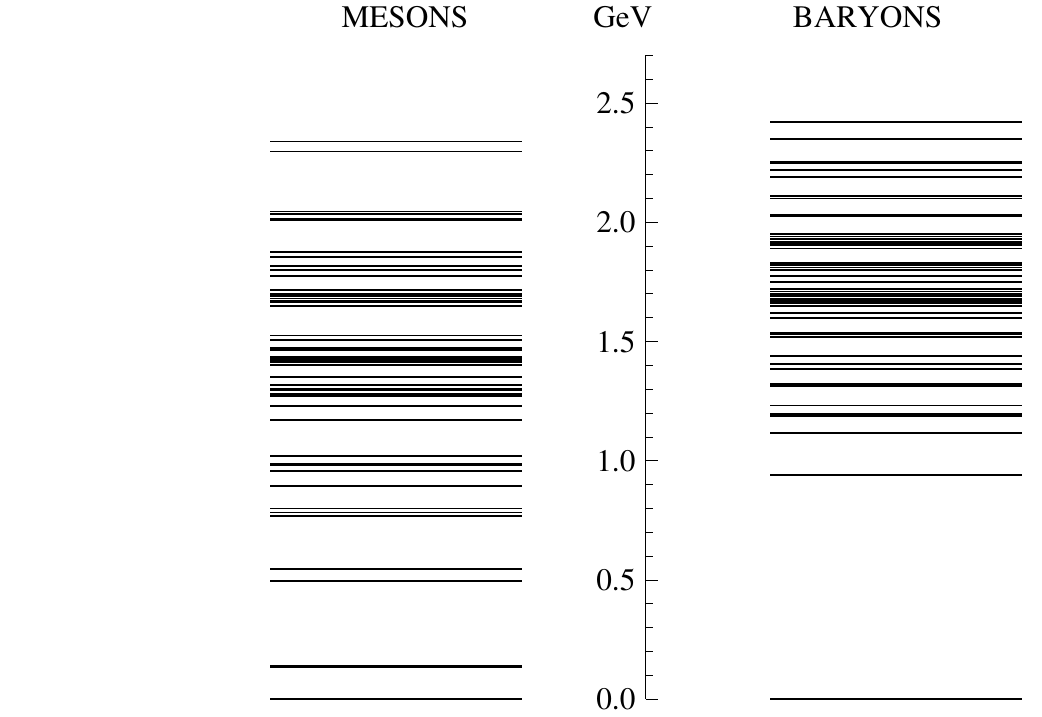} &
\includegraphics[width=0.43\textwidth]{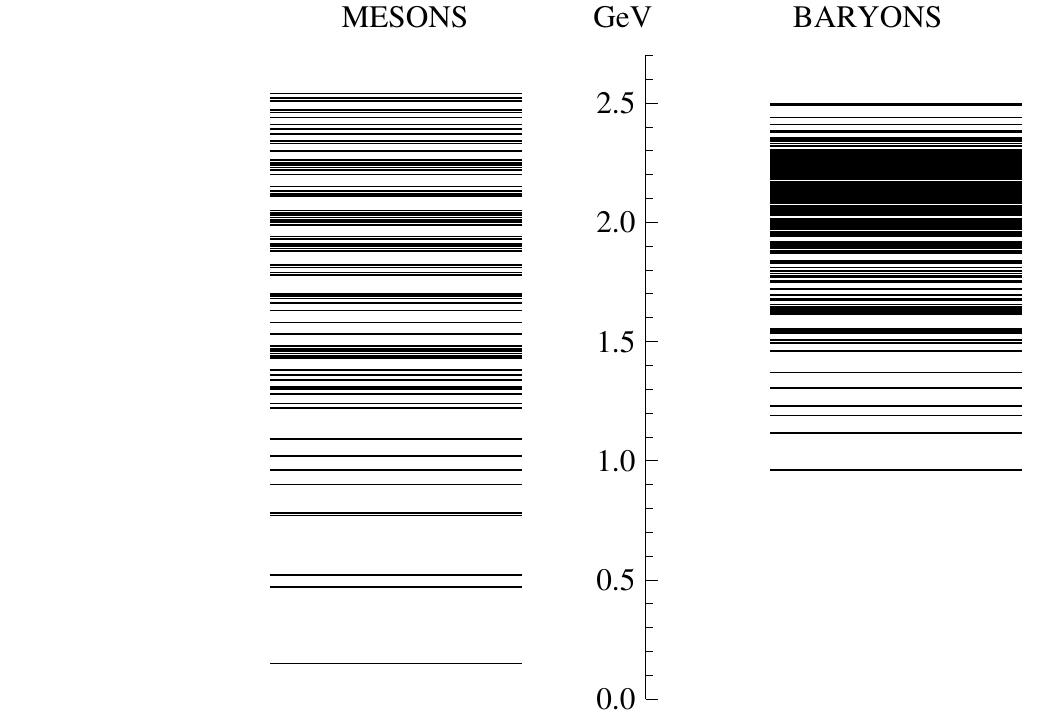} 
\end{tabular}
 \caption{\it Mesons and baryons spectrum made of $u$, $d$ and $s$ quarks from the PDG~\cite{Tanabashi:2018oca} (left panel) and from the relativized quark model~\cite{Godfrey:1985xj,Capstick:1986bm} (right panel). }
\label{fig:QCDspectrum}
\end{figure*}

The cumulative number of states is very useful for the characterization of the QCD spectrum. It is defined as the number of bound states below some mass~$M$, i.e.%
\begin{equation}
N(M) = \sum_i g_i  \, \Theta(M-M_i) \,,
\end{equation}
where $M_i$ is the mass of the $i$-th hadron, $g_i$ is the degeneracy, and $\Theta(x)$ is the step function, so that the density of states writes $\rho(M) = dN(M)/dM$. So far, the states listed in the PDG echo the standard quark model classification for mesons~$[q{\bar q}]$ and baryons~$[qqq]$, and the cumulative number will have contributions from any kind of states, i.e.
\begin{equation}
N(M) = M_{[q\bar{q}]}(M) + N_{[qqq]}(M) + N_{[q\bar{q}q\bar{q}]}(M) + \cdots \,.
\end{equation}
Then, it would be pertinent to consider also the spectrum of the RQM for hadrons, as it corresponds {\it by construction} to a solution of the quantum mechanical problem for both $q \bar q$-mesons and $qqq$-baryons. For color-singlet states, the $n$-parton Hamiltonian takes the form
\begin{equation}
H_n = \sum_{i=1}^n \sqrt{\vp_i^2+m_i^2} + \sum_{i< j}^n v_{ij}(\vr_{ij}) \,, \label{eq:H_RQM}
\end{equation}
where the two-body interactions take the form $v_{q \bar q}(r) =
-4\alpha_S/(3r) + \sigma r = (N_c - 1) v_{qq}(r)$. This Hamiltonian
for $n=2(3)$ describes the underlying dynamics of mesons(baryons). We
show in Fig.~\ref{fig:QCDspectrum} the hadron spectrum with the PDG
compilation (left) and the RQM spectrum (right). The comparison
clearly shows that there are further states in the RQM spectrum above
some scale $M > M_{\min}$ that may or may not be confirmed in the
future as mesons or hadrons, although they could also be exotic,
glueballs or hybrids.

\begin{figure*}[t]
\centering
 \begin{tabular}{c@{\hspace{4.5em}}c}
 \includegraphics[width=0.43\textwidth]{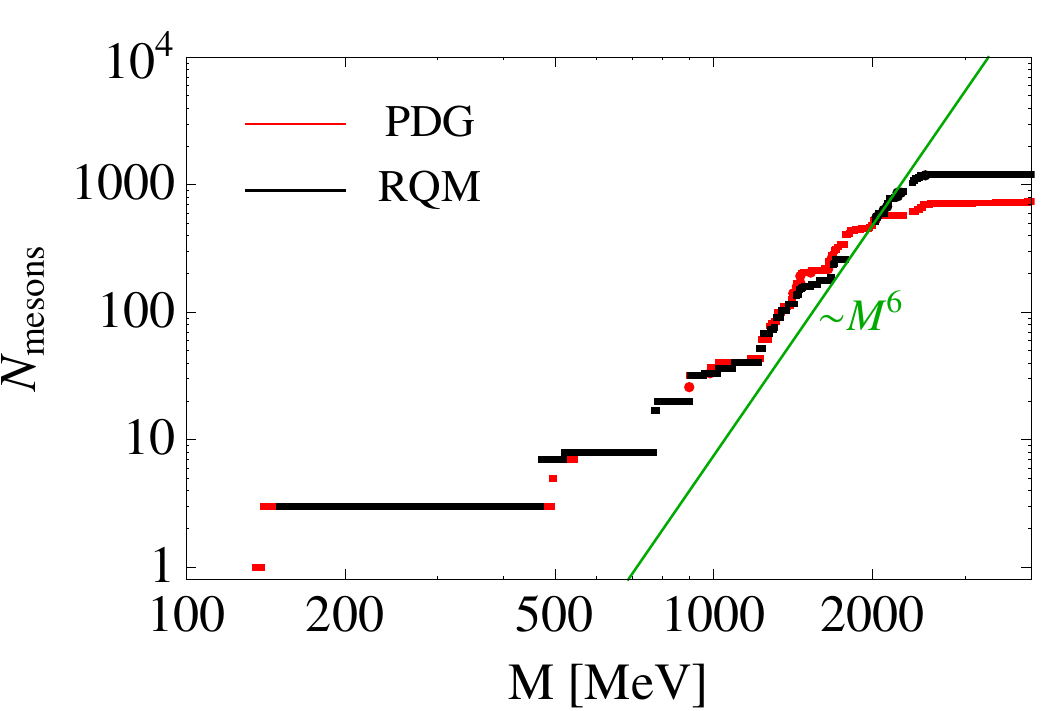} &
\includegraphics[width=0.43\textwidth]{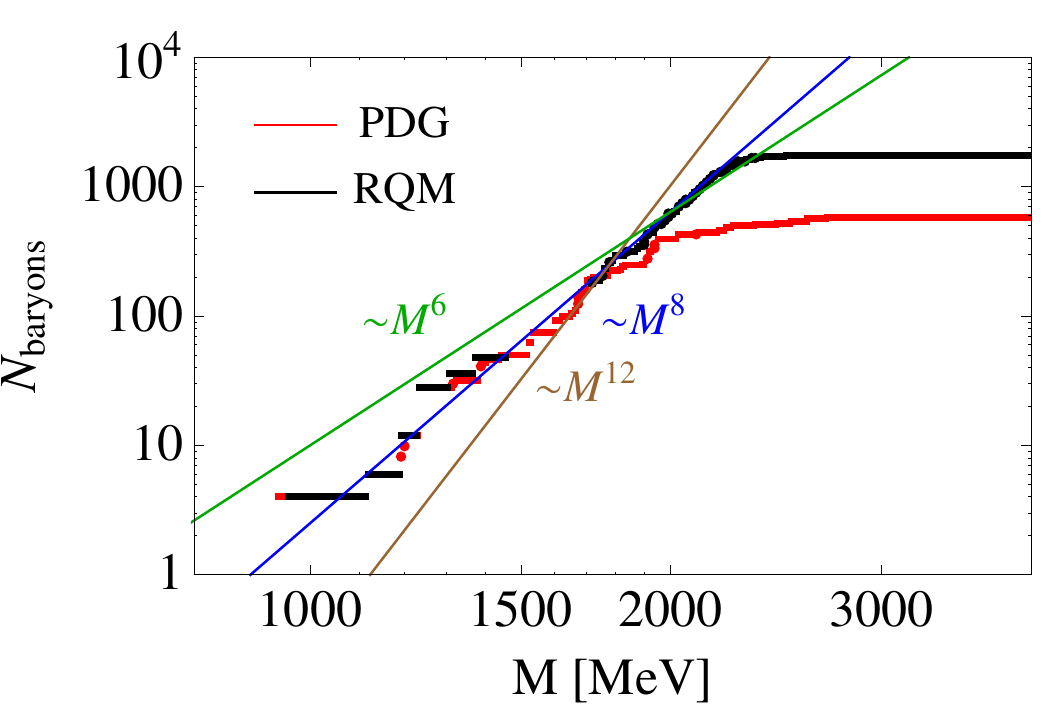} 
\end{tabular}
 \caption{\it Cumulative numbers for the PDG (lower curve) and the RQM
   (upper curve). We display in log-log scale the mesonic states (left panel), and the baryonic states (right panel).  }
\label{fig:nucum-meson-baryon}
\end{figure*}

A semiclassical expansion of the cumulative number of states~\cite{Caro:1994ht} can be used to study the high mass spectrum for systems where interactions are dominated by linearly rising potentials with a string tension~$\sigma$, in the range $M \gg \sqrt{\sigma}$. At leading order in the expansion, the cumulative number takes the form
\begin{equation}
\hspace{-22mm}
  N_n (M) \sim g_n \int \prod_{i=1}^n \frac{d^3 x_i d^3 p_i}{(2\pi)^3} \delta \left({\textstyle \sum_{i=1}^n \vx_i}\right) 
\delta \left({\textstyle \sum_{i=1}^n \vp_i}\right)  \Theta (M-H_n(p,x)) \sim   \left(\frac{M^2}{\sigma}\right)^{3 n-3} 
\end{equation}
where in the last estimate we have neglected the (color) Coulomb term of the potential. Then, one can predict that the large mass expansion of these contributions is 
\begin{equation}
N_{[q{\bar q}]} \sim M^6 \,, \quad N_{[qqq]} \sim M^{12} \,, \quad N_{[q{\bar q}q{\bar q}]} \sim M^{18} \,, \, \cdots  \,.
\end{equation}
This means that each kind of hadron dominates the function $N(M)$ at a
different scale. We display in Fig.~\ref{fig:nucum-meson-baryon} the
separate contributions of meson and baryon spectra for the PDG and
RQM. Note that while in the meson case the~$N_{[q{\bar q}]} \sim M^6$
behavior seems to conform with the asymptotic estimate, in the baryon
case much lower exponents, $M^{6}-M^8$, than the expected one are
identified. $M^6$ suggests a two-body dynamics, and we
take this feature as a hint that the $qqq$ excited spectrum
effectively conforms to a two body system of particles interacting
with a linearly growing potential. The consequences of this picture
will be analyzed in Sec.~\ref{sec:quark_diquark_model}.

Finally, let us mention that after adding all the contributions
$N_\lambda(M) \sim M^{q_\lambda}$, one obtains the conjectured behavior
of the Hagedorn spectrum~\cite{Arriola:2013jxa}
\begin{equation}
N_{\HRG}(M) \sim e^{M/T_H} \,.
\end{equation}
This exponential behavior would lead to a partition function
that becomes divergent at some finite value of the temperature,
i.e.
\begin{equation}
Z_{\HRG}  = \Tr \, e^{-H_{\HRG}/T}  \stackrel[T \to T_H^-]{\longrightarrow}{} \frac{A}{T_H - T} \,,
\end{equation}
where $T_H \approx 150 \MeV$ is the so-called Hagedorn temperature. The computation of the trace anomaly with the HRG model leads to a good description of the lattice data for $T \lesssim 0.8 \, T_c$ by using either the PDG or the RQM spectrum, see e.g. Ref.~\cite{Megias:2012hk}.

\section{Quark-diquark model for baryons}
\label{sec:quark_diquark_model}

In the quark model, such as the RQM~\cite{Capstick:1986bm}, baryons
are $[qqq]$ states where the interaction is given by a combination of
$\Delta$-like pairs of $qq$ interactions and a genuinely $Y$-like
$qqq$ interaction. An interesting possibility would be that the quarks
are distributed according to an isosceles triangle, leading to an
easily tractable class of models, the so-called quark-diquark ($qD$)
models~\cite{Ferretti:2011zz,Santopinto:2014opa,Gutierrez:2014qpa,Masjuan:2017fzu,RuizArriola:2017ggc}. In
these models, the baryons are assumed to be composed of a constituent
quark~$q$, and a constituent diquark~$D \equiv (qq)$, i.e. $B \equiv
[q(qq)]$. In its relativistic version the Hamiltonian writes
\begin{equation}
H_{qD} = \sqrt{\vp^2 + m_q^2} + \sqrt{\vp^2 + m_D^2} + V_{qD}(r) \,, \label{eq:HqD}
\end{equation}
corresponding to a two body problem. The aim of this section is to obtain the baryon spectrum within a simple realization of the $qD$ model. But before going to the phenomenological consequences of the model, we will show analytically that under very specific assumptions the $qD$ static interaction for heavy sources, $V_{qD}(r)$, coincides with the quark-antiquark potential, $V_{q\bar{q}}(r)$, up to an additive constant.

\subsection{Quark-diquark potential}
\label{subsec:quark_diquark_potential}

An operational way of placing static sources in a gauge theory is by introducing in the Euclidean formulation a local gauge rotation realized by the Polyakov loop, i.e.
\begin{equation}
\Omega(\vx) = \PP \, e^{\, i \int_0^\beta A_0(x) \, dx_0} \,,
\end{equation}
where $\PP$ indicates path ordering. The $q \bar q$ free energy is
given by the thermal expectation value (our color trace is {\it not
  normalized}, i.e.  $ \tr 1 = N_c$) \cite{Megias:2012kb}
\begin{equation}
e^{-F_{q \bar q} (r,T)/T}  =  \langle \tr \Omega(\vx_1)  \tr \Omega(\vx_2)^\dagger \rangle_T   \,,
\end{equation}
and the potential is obtained as the zero temperature limit of the free energy, i.e.~$V_{q \bar q}(r) = F_{q \bar q}(r,0)$. For baryons it is more relevant to study the~$qqq$ free energy, which is given by
\begin{equation}
e^{-F_{qqq} (\vx_1, \vx_2 , \vx_3 ,T)/T} = \langle \tr \Omega(\vx_1) \tr \Omega(\vx_2) \tr \Omega(\vx_3) \rangle_T \,.
\end{equation}
One can take in this expression the limit $\vx_3 \to \vx_2 $ to get the following result for the quark-diquark free energy 
\begin{equation}
e^{-F_{qD} (\vx_1, \vx_2 , T)/T} = \langle \tr \Omega(\vx_1) \tr (\Omega(\vx_2)^2)   \rangle_T  \,.
\end{equation} 
Note that this limit is singular since at very small distances the
interaction is dominated by one gluon exchange, $\sim 1/r $, and a
self-energy must be added. The renormalization of the composite
operator $\tr (\Omega(\vx_2)^2)$ yields an ambiguity in the form of an
additive constant from $F_{qqq}(\vx_1, \vx_2 , \vx_2 , T)$ to
$F_{qD}(\vx_1, \vx_2 , T)$. Finally, by using the Clebsch-Gordan
decomposition,~${\bf 3} \otimes {\bf 3}\otimes {\bf 3} = ( {\bf 3}
\otimes {\bf \bar{3}}) \oplus ( {\bf 3} \otimes {\bf 6} )$, one finds
that
\begin{equation}
e^{-F_{qD} (\vx_1, \vx_2 , T)/T} \equiv e^{-F_{q \bar q} (\vx_1, \vx_2, T )/T} + 
e^{-F_{3 \otimes 6} (\vx_1, \vx_2 , T )/T}  \simeq   e^{-F_{q \bar q} (\vx_1, \vx_2, T )/T} \,.
\end{equation}
In the last equality we have considered that the energy ${\bf 3}
\otimes {\bf 6}$ configuration is larger than that of the $q\bar{q}$
one. Hence, after taking the zero temperature limit, we find
\begin{equation}
V_{qD}(r) = V_{q\bar{q}}(r) + \textrm{const} \,,
\end{equation}
a property which is in marked agreement with recent lattice studies~\cite{Koma:2017hcm} (see also Refs.~\cite{Greenberg:1981xn,Savage:1990di}), and that will be used in the following.

\subsection{The model}
\label{subsec:model}

Based on the considerations above, we can assume for the $qD$ potential~\cite{Megias:2018haz}
\begin{equation}
V_{qD}(r) = - \frac{\tau}{r} + \sigma r + \mu \,,  \label{eq:VqD}
\end{equation}
with $\tau = \pi/12$ and $\sigma = (0.42\GeV)^2$. The parameters of the kinetic terms in the Hamiltonian of Eq.~(\ref{eq:HqD}) are controlled by: i) the constituent quark mass, $m_\cons$, and ii) the current quark mass for the strange quark, $\hat{m}_s$; in the following way
\begin{equation}
m_{u,d} = m_\cons   \,, \qquad m_s = m_\cons + \hat{m}_s \,.
\end{equation}
In addition, we can distinguish between two kinds of diquarks: scalar $D \equiv [q_1 q_2]$, and axial vector $D_{AV} \equiv \{q_1 q_2\}$; so that in the following we will consider the natural choice
\begin{equation}
m_{D, \nns} = 2 m_{\cons} \,, \qquad m_{D_{AV}, \nns} = m_{D, \nns} + \Delta m_D \,, \label{eq:mDsAV}
\end{equation}
where the subindex $\textrm{{ns}}$ refers to diquarks with non-strange
quarks. Some studies in QCD indicate a mass difference between these
diquarks of $\Delta m_D \simeq 0.21 \GeV$~\cite{Jaffe:2004ph}, a value
that will be adopted in the following. On the other hand, the breaking
of flavor $\SU(3)$ for diquarks will be modeled as
\begin{equation}
m_X = m_{X, \nns }  + n_s  \hat{m}_s  \,, \qquad X = D, D_{AV} \,,
\label{eq:mDsAV2}
\end{equation}
where $n_s$ is the number of $s$ quarks in the diquark. Under these assumptions, the only parameters of the model that remain free are~$m_\cons$, $\hat{m}_s$ and $\mu$. We summarize in Table~\ref{tab:degeneracy} the degeneracies of the states predicted by the model, by distinguishing between their electric charges. The quantum numbers of these states are $Q$, $B = 1$ and $S = -n_s$. 
\begin{table}[htb!]
\caption{Spin-isospin degeneracies of the baryonic states within the
  quark-diquark model. $n$ represents the light flavors $u$ and $d$, while $s$ is the  strange quark.}
\label{tab:degeneracy}
\vspace{0.2cm}
\centering
\begin{tabular}{||c|c|c|c|c|c||}
\hline\hline
Baryon & Total deg. & $Q= -1$ & $Q=0$ & $Q=1$ & $Q=2$  \\ \hline
$[nn]n$     &   4    &  -  &   2  &  2  &  -  \\   \hline
$\{nn\}n$   &  36    &  6  &  12  & 12  &  6   \\  \hline
$[nn]s$     &   2    &  -  &   2  &  -  &  -   \\  \hline
$\{nn\}s$   &  18    &  6  &   6  &  6  &  -   \\  \hline
$[ns]n$     &   8    &  2  &   4  &  2  &  -   \\  \hline
$\{ns\}n$   &  24    &  6  &  12  &  6  &  -   \\  \hline
$[ns]s$     &   4    &  2  &   2  &  -  &  -   \\  \hline
$\{ns\}s$   &  12    &  6  &   6  &  -  &  -   \\  \hline
$\{ss\}n$   &  12    &  6  &   6  &  -  &  -   \\  \hline
$\{ss\}s$   &   6    &  6  &   -  &  -  &  -   \\ 
\hline\hline
\end{tabular}
\end{table}

\subsection{Baryon spectrum}
\label{subsec:baryon_spectrum}

\begin{figure*}[t]
 \includegraphics[width=0.43\textwidth]{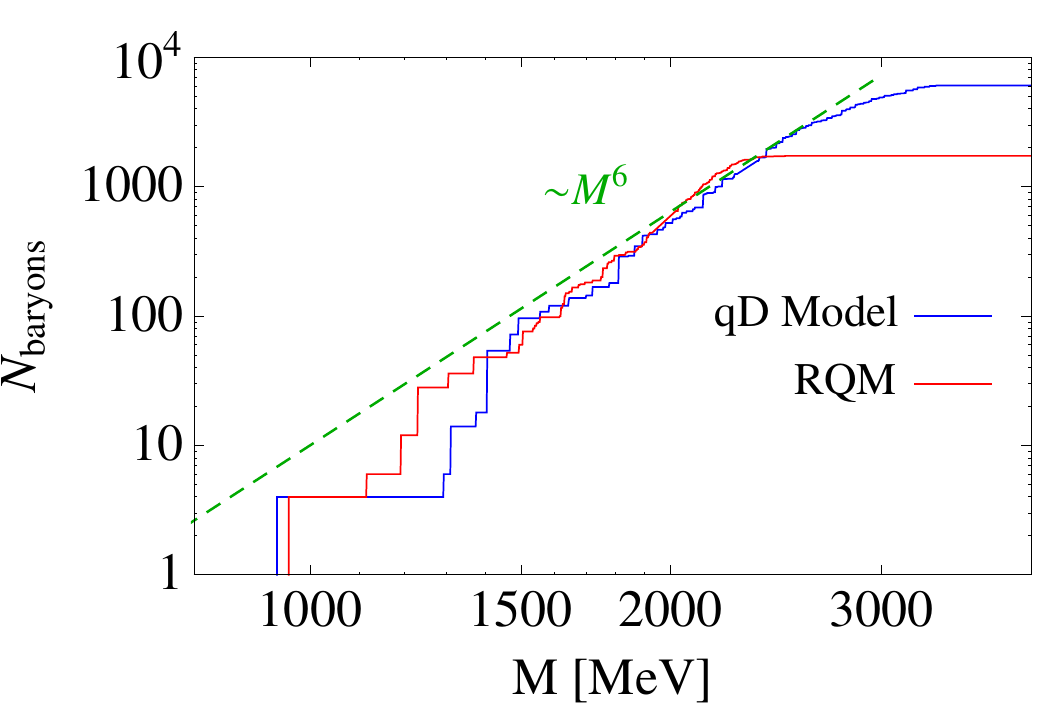} 
\hspace{-1cm} \begin{minipage}{23pc}
\vspace{-3.3cm} \caption{\it Cumulative number for the baryon spectrum
  as a function of the baryon mass. We compare the $qD$ model of
  Sec.~\ref{sec:quark_diquark_model}, and the
  RQM~\cite{Capstick:1986bm}.}
\label{fig:N_chi_qD}
\end{minipage}
\end{figure*}

The baryon spectrum of the $qD$ model can be obtained by diagonalizing the Hamiltonian of Eq.~(\ref{eq:HqD}). We will consider a variational procedure by using as basis the 3-dimensional isotropic harmonic oscillator (IHO), with normalized wave functions of the form
\begin{equation}
R_{n\ell}(r) = \sqrt{ \frac{(n-1)! \, 2^{\ell+n+1}}{ \sqrt{\pi} b^3
    (2\ell + 2(n-1) + 1)!!}  }  L_{n-1}^{\ell + \frac{1}{2}} (r^2/b^2)
\left(\frac{r}{b} \right)^\ell
e^{-\frac{r^2}{2b^2}}
\,,
\label{eq:Rnl}
\end{equation}
where $L_{n-1}^{\ell + \frac{1}{2}}(x)$ are the generalized Laguerre polynomials. The matrix elements of the Hamiltonian are then obtained from
\begin{eqnarray}
 \langle n\ell | H_{qD} | n^\prime \ell \rangle &=& \int_0^\infty dp
 \, \hat{u}_{n\ell}^\ast(p) \hat{u}_{n^\prime \ell}(p) \left[
   \sqrt{p^2 + m_q^2 } + \sqrt{p^2 + m_D^2} \right] \nonumber \\
&& + \int_0^\infty dr \, u_{n\ell}^\ast(r) u_{n^\prime \ell}(r) V_{qD}(r) \,,  \label{eq:matrix_element_H}
\end{eqnarray}
with $u_{n\ell}(r)$ and $\hat{u}_{n\ell}(p)$ the reduced wave functions of the IHO in position and momentum space, respectively. The value of the parameter $b$ is fixed to minimize the energy levels for each of the multiplets in Table~\ref{tab:degeneracy}, being the typical values of this parameter in the range $0.55 \fm \lesssim b \lesssim 0.65 \fm$. We get with this procedure the spectrum of baryons that is shown in Fig.~\ref{fig:N_chi_qD}. We have considered as a convenient choice of the parameters of the model, the values
\begin{equation}
\hspace{-15mm}
m_{D, \nns } = 0.6 \GeV \,, \quad m_{u,d} = 0.3 \GeV \,, \quad \hat{m}_s = 0.10 \GeV \,, \quad \mu = -0.459 \GeV \,. \label{eq:param}
\end{equation}
These values are motivated by the fits of the lattice data for the baryonic susceptibilities that will be presented in Sec.~\ref{subsec:Baryonic_sus_qD}. It is remarkable that below $M < 2400 \MeV$ the quark-diquark spectrum is in good agreement with the RQM spectrum. Note that the growth of the cumulative number is $N_{[qD]} \sim M^{6}$, as expected for a two body problem. Since the quark-diquark picture is reliable only for excited states, in the following we will use the empirical value of the mass of the nucleon, $M_n = 938 \MeV$, and apply the~$qD$ model only for the other baryons.

\section{Fluctuations of conserved charges and baryonic susceptibilities}
\label{sec:fluctuations}

 The EoS of QCD is sensitive to both mesonic and baryonic states, and
 the thermodynamic separation of these degrees of freedom cannot be
 done at the QCD level. However, there is a number of thermal
 observables that can be used to study the spectrum of QCD by
 distinguishing between different flavor sectors; these are the
 fluctuations~\cite{Bazavov:2012jq} and
 correlations~\cite{Megias:2019osk} of conserved charges. Beyond these
 hadron spectrum considerations, recent lattice studies of
 fluctuations suggest that they are sensitive probes of deconfinement
 in realistic physical situations in which quarks cannot be considered
 as heavy sources (see e.g. Ref.~\cite{Ding:2015ona} for a recent
 review). We will study in this section the baryonic susceptibilities
 within the HRG approach by using the $qD$ spectrum obtained in
 Sec.~\ref{sec:quark_diquark_model}. A comparison with the lattice
 data for the fluctuations will allow to extract information about the
 values of the parameters of the model.

\subsection{Fluctuations of conserved charges}

Conserved charges $[Q_a,H]=0$ play a fundamental role in the thermodynamics of QCD. When considering the $(uds)$ flavor sector of QCD, the only conserved charges are the electric charge $Q$, the baryon number $B$, and the strangeness~$S$. Their thermal expectation values present statistical fluctuations, that are usually computed from the grand-canonical partition function~\cite{Asakawa:2015ybt}
\begin{equation}
Z_{\QCD} = \textrm{Tr} \exp\bigg[ - \Big( H_{\QCD} - \sum_a \mu_a Q_a \Big)/T \bigg] \,.
\end{equation}
By considering the differentiation of the thermodynamical potential $\Omega = -T \log Z_{\QCD}$ with respect to the chemical potential, one finds the thermal expectation values and the susceptibilities of the charges
\begin{equation}
\hspace{-17mm}
  \langle Q_a \rangle_T  = -\frac{\partial \Omega}{\partial\mu_a} \Bigg|_{\mu_a = 0}\,, \quad \chi_{ab}(T) \equiv  \frac{1}{V T^3}\langle \Delta Q_a \Delta Q_b \rangle_T =  - \frac{1}{VT^2} \frac{\partial^2 \Omega}{\partial\mu_a \partial\mu_b}  \Bigg|_{\mu_a = 0 = \mu_b}  \,, \label{eq:susceptibilities}
\end{equation}
where $\Delta Q_a = Q_a - \langle Q_a \rangle_T$, and $Q_a \in \{Q,B,S\}$. We have used that $\langle Q_a \rangle_T = 0$ in absence of chemical potentials. Note that the fluctuations can also be computed in the quark-flavor basis, $Q_a \in \{ u, d, s\}$, where $u$, $d$ and $s$ is the number of up, down and strange quarks. In this basis
\begin{equation}
B = \frac{1}{3}(u+d+s) \,, \quad Q = \frac{1}{3}(2u-d-s) \,, \quad S= -s \,.
\end{equation}

\subsection{Fluctuations within the HRG approach}
\label{subsec:fluctuations_HRG}

While in the deconfined phase of QCD the quarks and gluons are
liberated to form a plasma, in the confined/chiral symmetry broken
phase the relevant degrees of freedom are bound states of quarks and
gluons, i.e. hadrons and possibly exotic states. This means that it
should be expected that physical quantities in this phase admit a
representation in terms of hadronic states. This is the idea of the
hadron resonance gas (HRG) model which describes the equation of state of QCD in terms of a free gas of hadrons~\cite{Hagedorn:1984hz,Huovinen:2009yb,Megias:2009mp},
\begin{equation}
\log Z_{\HRG} = - V \int \frac{d^3p}{(2\pi)^3} \sum_{i \in {\rm Hadrons}} \!\! \zeta_i \, g_i \log\left( 1 - \zeta_i \, e^{-\left( E_{p,i} - \sum_a \mu_a q_i^a  \right)/T} \right)  \,, \label{eq:ZHRG}
\end{equation}
where $E_{p,i} = \sqrt{p^2 + M_i^2}$, $\zeta_i = \pm 1$ for bosons/fermions, and $q^a_i \in \{ Q_i, B_i , S_i\}$ is the charge of $i$th-hadron for symmetry~$a$. Within the HRG approach, the charges are carried by various species of hadrons, so that~$Q_a =  \sum_{i \in {\rm Hadrons}} q^a_i N_i$, where $N_i$ is the number of hadrons of type~$i$. The susceptibilities within the HRG model can be obtained by just applying Eq.~(\ref{eq:susceptibilities}) to the partition function of Eq.~(\ref{eq:ZHRG}). This leads to the following result
\begin{equation}
\chi_{ab}(T) =   
 \frac{1}{2\pi^2}  \sum_{i \in {\rm Hadrons}} \!\! g_i \, q_i^a \, q_i^b \sum_{n=1}^\infty  \zeta_i^{n+1} \frac{M_i^2}{T^2} K_2\left( \frac{nM_i}{T} \right)  \,, \label{eq:chi_HRGM}
\end{equation}
where $K_2(z)$ is the Bessel function of the second kind. Some previous studies of the second order fluctuations within the HRG approach lead to a good description of lattice data for $T \lesssim 160$ MeV, as expected from a description based on hadronic degrees of freedom and as such valid only in the confined phase of QCD, see e.g. Refs.~\cite{RuizArriola:2016qpb,Megias:2019osk}. In particular, fluctuations have been proposed as a diagnostic tool to study missing states in the different charge sectors by considering a comparison of the HRG approach with lattice data~\cite{RuizArriola:2016qpb}. Note that Eq.~(\ref{eq:chi_HRGM}) predicts the asymptotic behavior
\begin{equation}
\chi_{ab}(T)   \stackbin[T \to 0]{}{\sim} e^{-M_{0}^{ab}/T} \,,  \label{eq:Chi_lowT}
\end{equation}
\begin{figure*}[t]
 \includegraphics[width=0.43\textwidth]{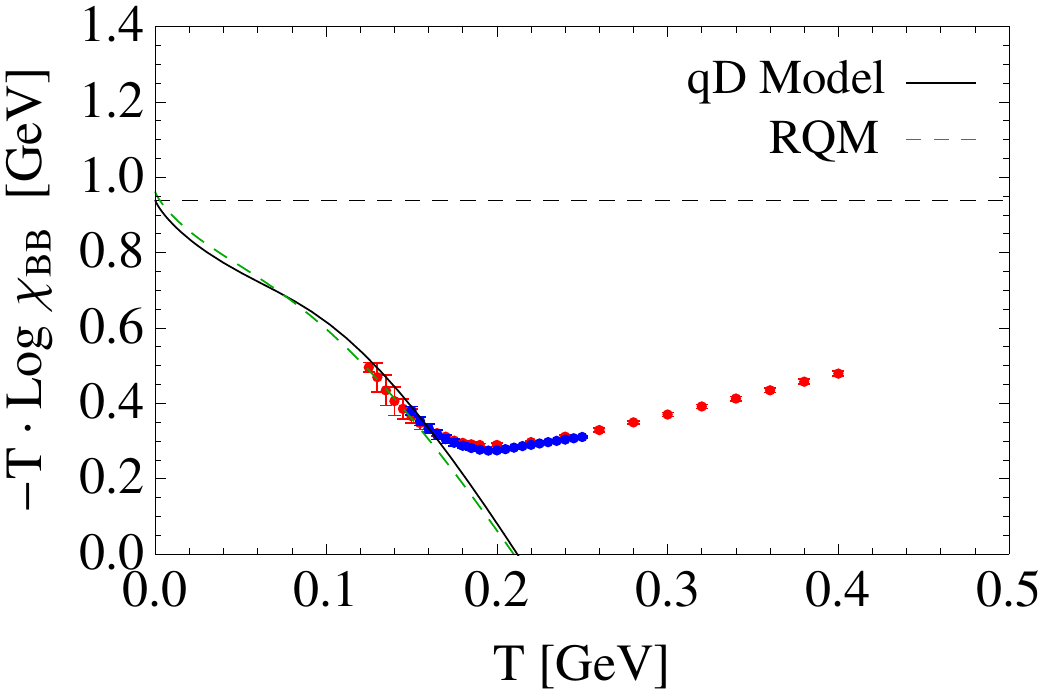} 
\hspace{-1cm}\begin{minipage}{23pc}
\vspace{-4.3cm} \caption{\it Plot of $-T \log |\chi_{BB}|$ as a
  function of temperature. We display as dots the lattice data from
  Refs.~\cite{Borsanyi:2011sw} (red) and \cite{Bazavov:2012jq}
  (blue). We also display the HRG model results including the spectrum of the $qD$ model (solid black), and the RQM spectrum~\cite{Capstick:1986bm} (dashed green).}
\label{fig:LogChi1}
\end{minipage}
\end{figure*}
where $M_0^{ab}$ is the mass of the lowest-lying state in the spectrum
with quantum numbers~$a$ and $b$. For the baryonic susceptibilities,
these states are $M_{0}^{BB} = M_{0}^{BQ} = M_p$ the proton mass, and
$M_{0}^{BS} = M_{\Lambda^0}$ the $\Lambda^0$ baryon mass. This
observation, which goes beyond the HRG model as the lightest hadron
saturates the QCD partition function in each sector at low enough
temperature, makes it appealing to plot the lattice data for the
susceptibilities in logarithmic scale. The corresponding plot for
$\chi_{BB}$ is displayed in Fig.~\ref{fig:LogChi1}, and compared with
the HRG approach including: i) the quark-diquark model spectrum
computed in Sec.~\ref{sec:quark_diquark_model}, and ii) the RQM baryon
spectrum~\cite{Capstick:1986bm}.  The horizontal dashed line
represents the value of the lowest-lying state contributing to the
fluctuation (the proton), cf. Eq.~(\ref{eq:Chi_lowT}). In practice we
find that at the lowest available temperatures, the contribution of excited states becomes individually small but collectively important. This is a typical problem in intermediate temperature analyses.

\subsection{Baryonic susceptibilities with the $qD$ model spectrum}
\label{subsec:Baryonic_sus_qD}

From the spectrum of the $qD$ model, we can obtain the baryonic susceptibilities by using the HRG approach given by Eq.~(\ref{eq:chi_HRGM}). Since our goal is to reproduce the lowest temperature values of the lattice results for these quantities, we have chosen to minimize the function
\begin{equation}
\bar\chi^2 = \bar\chi_{BB}^2 +   \bar\chi_{BQ}^2 +  \bar\chi_{BS}^2 \,, \quad \textrm{where} \quad
\bar\chi_{ab}^2 = \sum_{j=1}^{j_{\max}} \frac{\left( \chi_{ab}^{\lat}(T_j) -
  \chi_{ab}^\HRG(T_j)\right)^2}{(\Delta \chi_{ab}^{\lat}(T_j))^2} 
\,, \label{eq:barchi2}
\end{equation}
\begin{figure*}[t]
\centering
 \begin{tabular}{c@{\hspace{3.5em}}c}
 \includegraphics[width=0.43\textwidth]{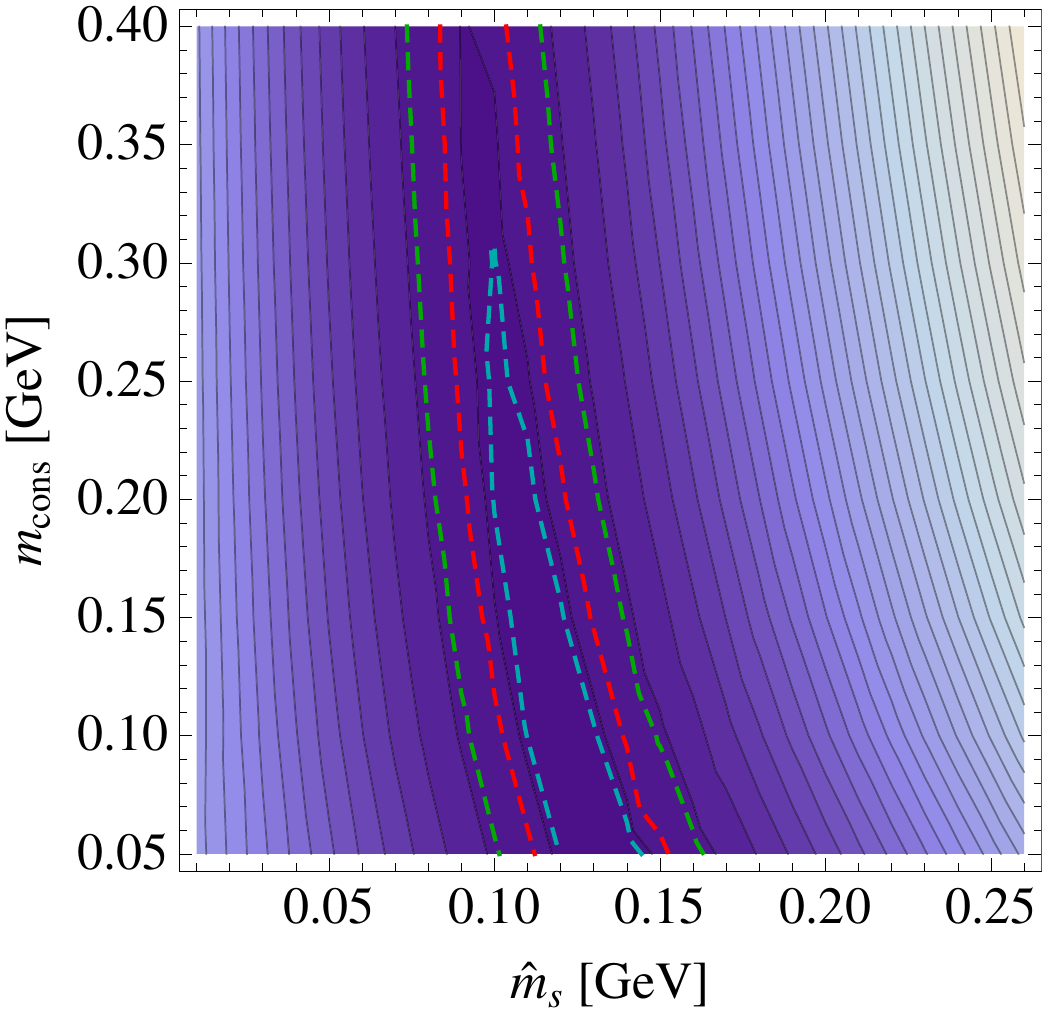} &
\includegraphics[width=0.43\textwidth]{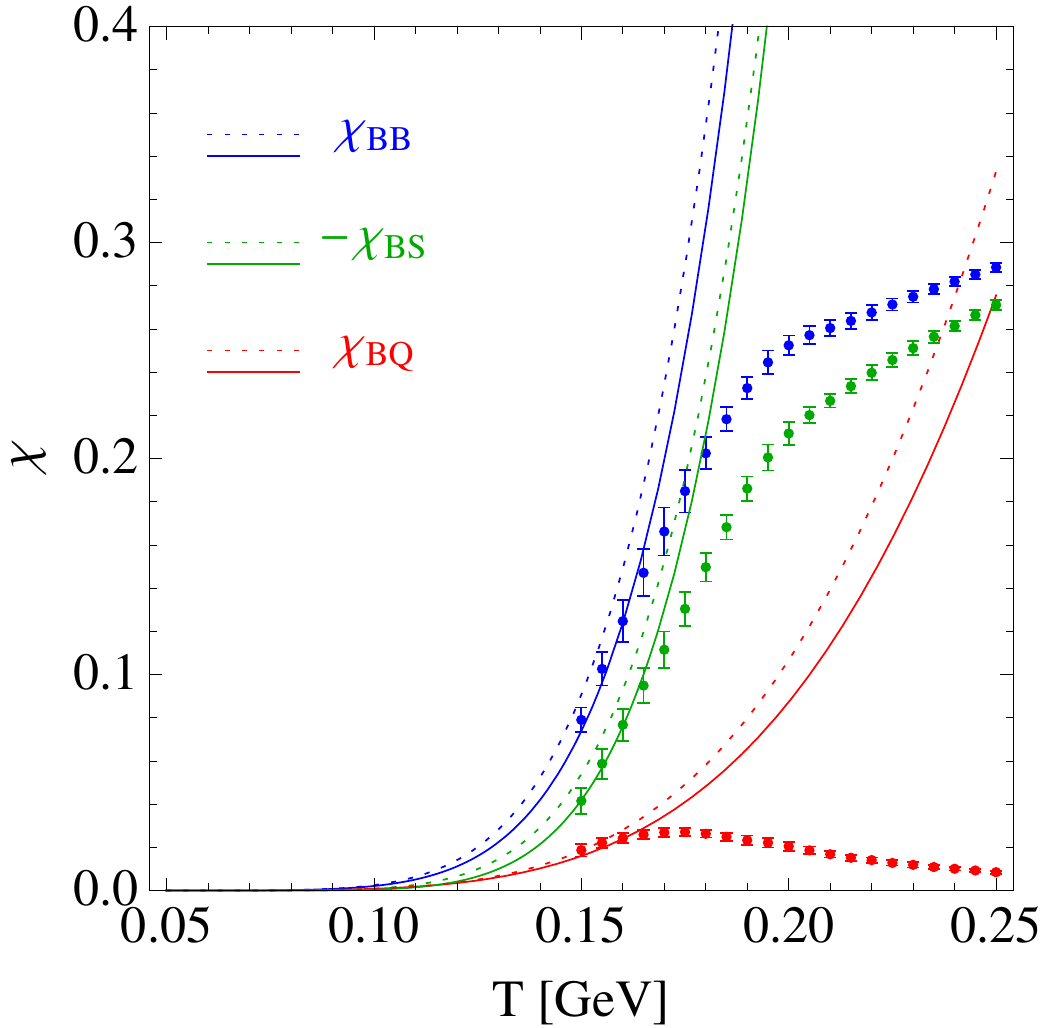} 
\end{tabular}
 \caption{\it Left panel: $\bar\chi^2/\nu$ in the plane $(\hat{m}_s,
   m_{\cons})$ from a fit to the lattice data of the baryonic
   fluctuations from~\cite{Bazavov:2012jq} with $T \le 165\MeV$. The
   dashed lines correspond to $\bar\chi^2/\nu = 0.77$ (curve
   delimiting a smaller area), $\bar\chi^2/\nu = 1$, and
   $\bar\chi^2/\nu = 1+\sqrt{2/\nu}$ (curve delimiting a larger
   area). Right panel: Baryonic susceptibilities from the
   quark-diquark model (solid) compared to the lattice data of
   Ref.~\cite{Bazavov:2012jq}.  We display also as dotted lines the
   results from the spectrum of the RQM~\cite{Capstick:1986bm}. For
   the quark-diquark model we have used the parameters in
   Eq.~(\ref{eq:param}).  }
\label{fig:ms_mcons}
\end{figure*}
and $j_{\max}$ is the number of data points used in the fits. As
explained in Sec.~\ref{subsec:model}, the model contains three free
parameters. We show in the left panel of Fig.~\ref{fig:ms_mcons} a
plot of $\bar\chi^2/\nu$, where $\nu$ is the number of degrees of
freedom, in the plane $(\hat{m}_s, m_{\cons})$.~\footnote{Whenever we
  provide a value for $\bar\chi^2/\nu$, it should be understood that
  this function is already minimized with respect to the
  parameter~$\mu$ in Eq.~(\ref{eq:VqD}).}  One can observe from this figure
that the current quark mass for the strange quark takes a value
compatible with the PDG, i.e. $80 \MeV \lesssim \hat{m_s} \lesssim 120
\MeV$, while the constituent quark mass is in the range $100 \MeV
\lesssim m_{\cons}\lesssim 400 \MeV$. Based on these results, we
propose the choice of the parameters of the model already presented in
Eq.~(\ref{eq:param}). Using them, we obtain the results for the
baryonic susceptibilities that are displayed in the right panel of
Fig.~\ref{fig:ms_mcons}.

In some previous works, the Coulombic term of the $qD$ potential is regularized at the origin $(r=0)$ by adding a term of the form $\Delta V_{qQ}(r) = \tau e^{-\kappa r} /r$ in Eq.~(\ref{eq:VqD}), see e.g. Refs.~\cite{Santopinto:2014opa,DeSanctis:2014ria}.  To study the effects of this term, we have considered the following values for the regularization parameter: $\kappa = (10 \fm^{-1},  40 \fm^{-1}, 80 \fm^{-1})$. This range includes most of the values considered in the literature. Then we find that this correction increases the lowest states of the baryon spectrum by $\Delta m_n / m_n \sim~(0.7 \% \,, 0.05 \% \,,  0.015 \%)$ for $\kappa = (10 \fm^{-1},  40 \fm^{-1}, 80 \fm^{-1})$, being this increase less important for heavier states. These effects in the spectrum induce small corrections in the baryonic susceptibilities that tend to decrease their absolute values by $|\delta \chi_{ab}^{\HRG}|/|\chi_{ab}^{\HRG}| \sim~(2 \% ,  0.2\% , 0.02\% )$ respectively, in the temperature regime $T \gtrsim 0.1 \GeV$, which is the relevant one for the fits performed in the present work.~\footnote{We find that the relative corrections in the susceptibilities induced by $\Delta V_{qD}(r)$ are in general larger for lower temperatures. However, this is an effect of the smallness of the susceptibilities in this regime of temperatures (specially for $T \lesssim 0.1 \, \GeV$), and in fact the absolute corrections in the susceptibilities tend to decrease at lower temperatures.}

\subsection{Baryonic susceptibilities of higher order}
\label{subsec:higher_order}

\begin{figure*}
\begin{center}
 \begin{tabular}{c@{\hspace{0.5em}}c}
 \includegraphics[width=0.43\textwidth]{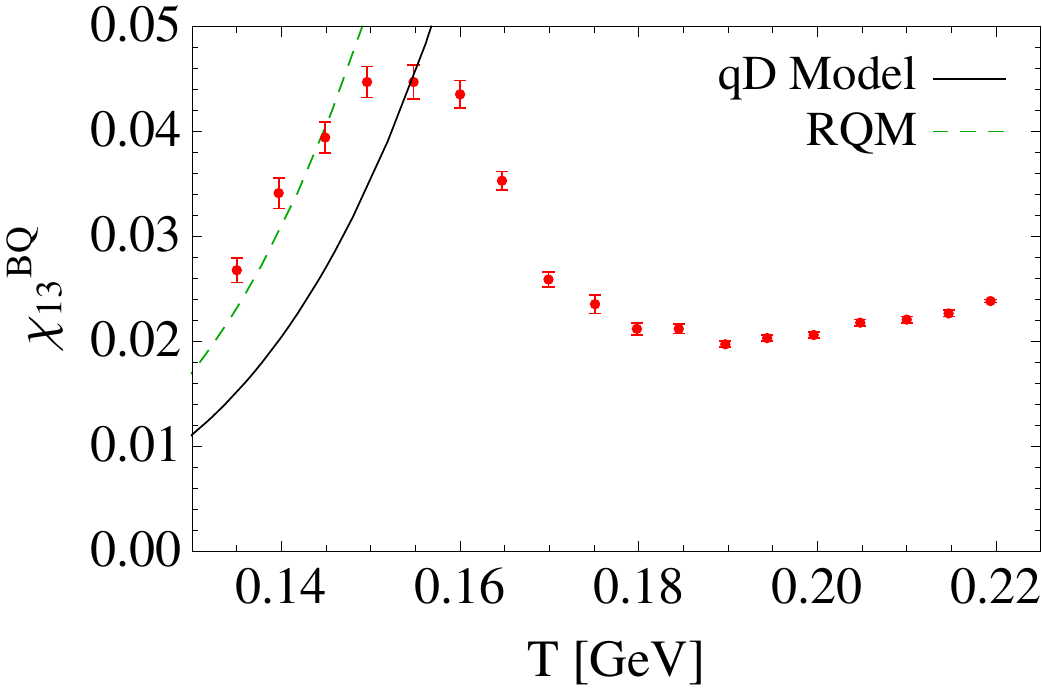} & \hspace{1cm}
 \includegraphics[width=0.43\textwidth]{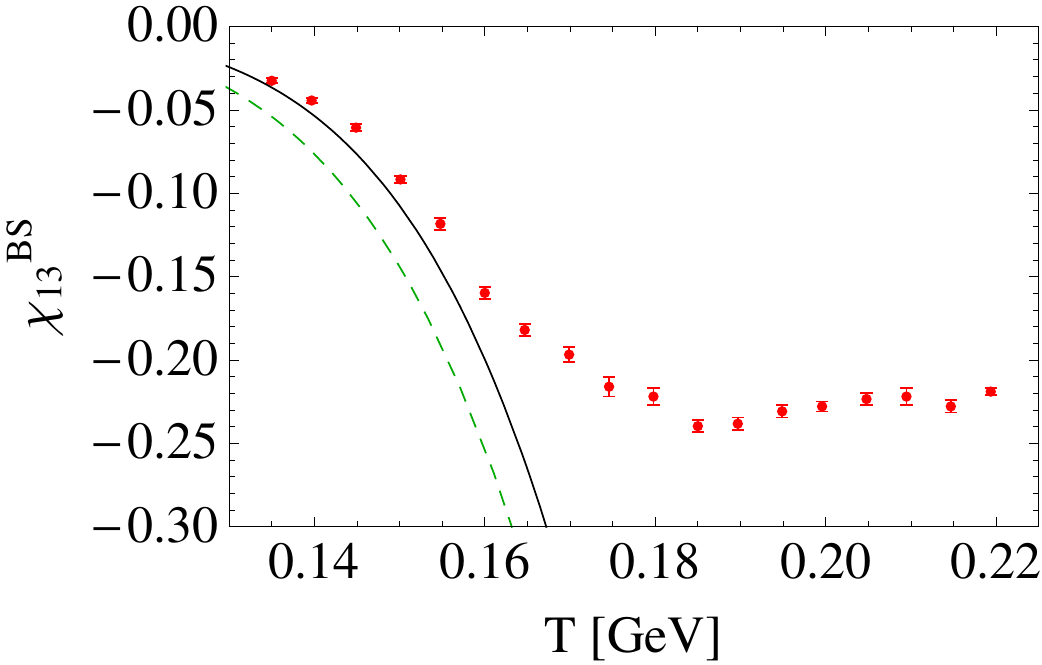}  \\
 \includegraphics[width=0.43\textwidth]{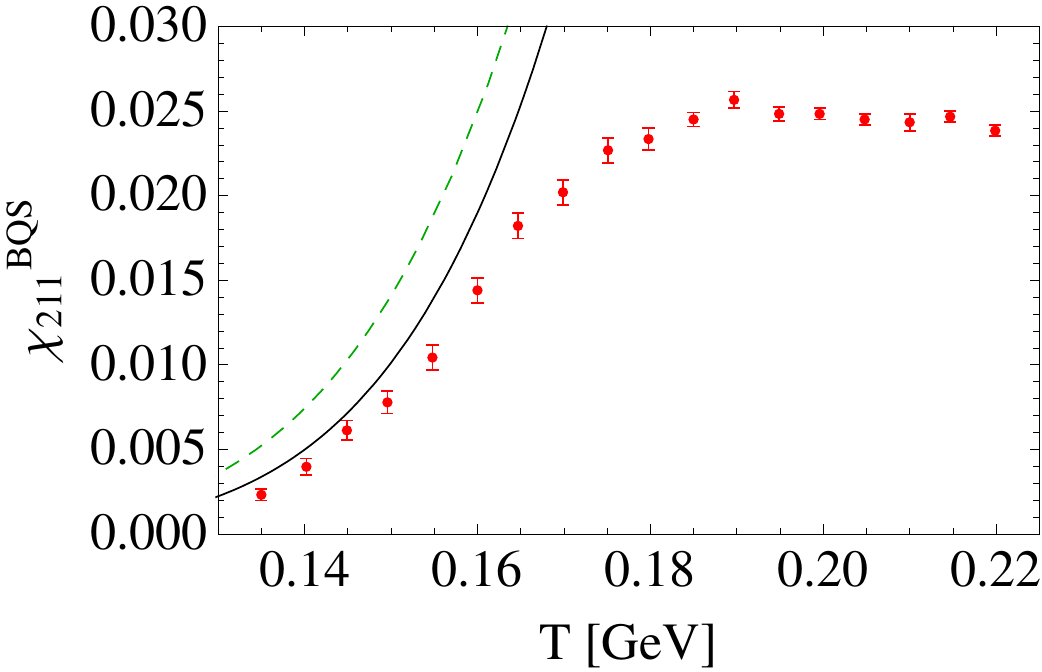} & \hspace{1cm}
 \includegraphics[width=0.43\textwidth]{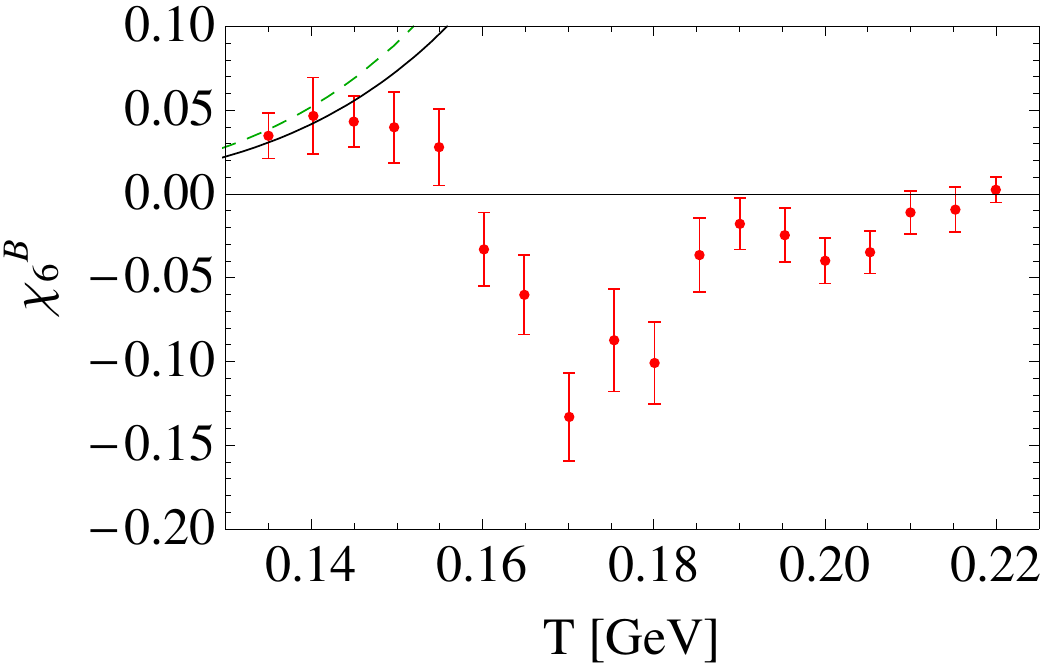}  \\
 \includegraphics[width=0.43\textwidth]{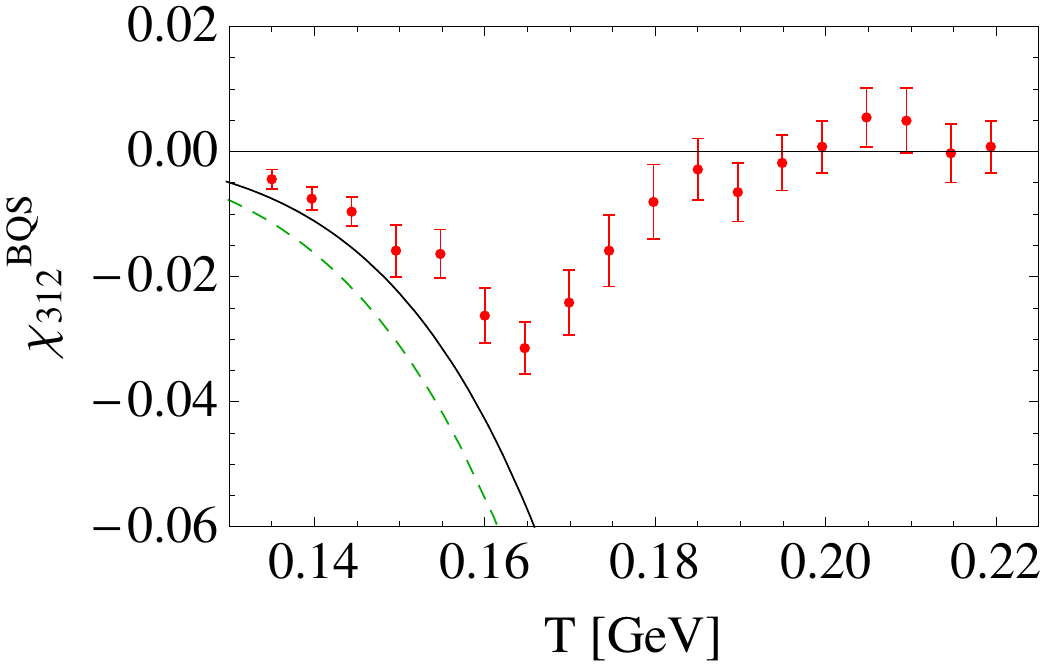} & \hspace{1cm}
 \includegraphics[width=0.43\textwidth]{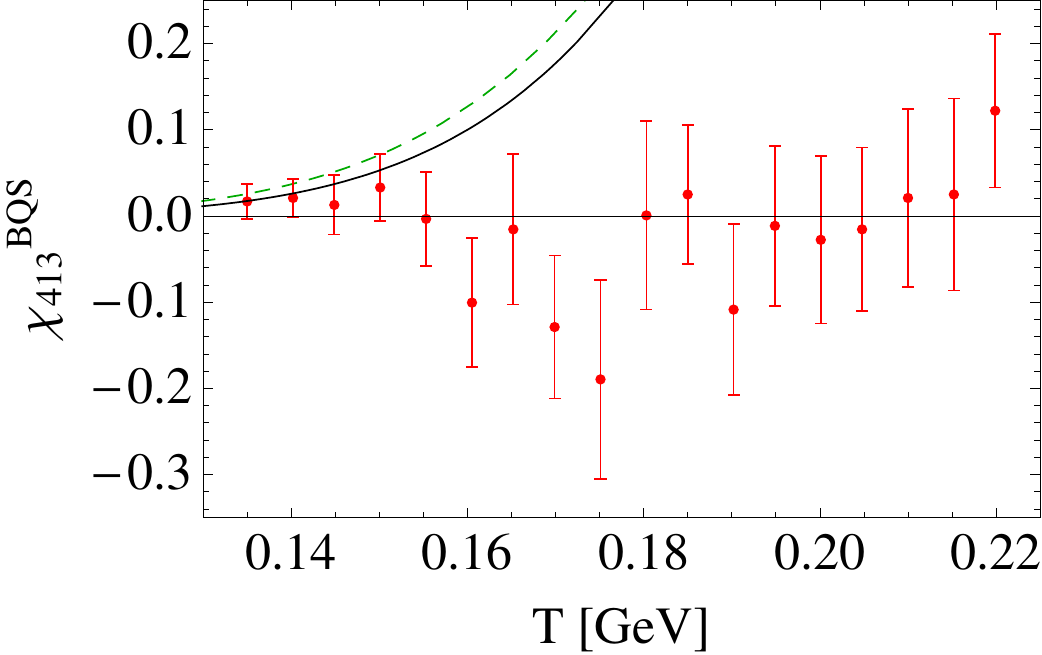}  \\
\end{tabular}
\end{center}
\vspace{-0.5cm}
 \caption{Baryonic susceptibilities of fourth order ($\chi_{13}^{BQ}$,
   $\chi_{13}^{BS}$, $\chi_{211}^{BQS}$), sixth order ($\chi_6^B$,
   $\chi_{312}^{BQS}$) and eighth order ($\chi_{413}^{BQS}$) from the
   quark-diquark model (solid black). We have used the parameters in
   Eq.~(\ref{eq:param}). We display as dots the lattice data of
   Ref.~\cite{Borsanyi:2018grb}, and as dashed (green) lines the
   results by using the HRG model with the baryonic spectrum of the
   RQM~\cite{Capstick:1986bm}.}
\label{fig:chi468}
\end{figure*}

Higher order fluctuations can be obtained within the HRG model by taking higher derivatives of the thermodynamical potential. The corresponding susceptibility of order $(p,q,r)$ in charges $(B,Q,S)$ reads
\begin{eqnarray}
\chi^{BQS}_{pqr}(T) &=  \frac{\partial^{p+q+r}\left(p/T^4 \right)}{ (\partial\hat\mu_B)^p (\partial\hat\mu_Q)^p (\partial\hat\mu_S)^r } \Bigg|_{\hat\mu_a = 0}  \nonumber \\
&=  \frac{1}{2\pi^2}  \sum_{i \in {\rm Hadrons}} 
g_i \, B_i^p  Q_i^q  S_i^r  \; \sum_{n=1}^\infty \zeta_i^{n+1} 
n^{p+q+r-2} \,\frac{M_i^2}{T^2} K_2\!\left( \frac{nM_i}{T} \right)  \,, \label{eq:chi_HRGMhigher}
\end{eqnarray}
where $p = -\Omega/V$ is the pressure, and $\hat\mu_a = \mu_a/T$.~\footnote{With the notation of Eq.~(\ref{eq:chi_HRGMhigher}), the second order susceptibilities studied in Secs.~\ref{sec:fluctuations} and \ref{sec:semiclassical} are $\chi_{BB} \equiv \chi_{200}^{BQS}$,  $\chi_{BQ} \equiv \chi_{110}^{BQS}$ and  $\chi_{BS} \equiv \chi_{101}^{BQS}$.} We have studied within the HRG approach some of the baryonic susceptibilities of fourth, sixth and eighth  order, i.e. $p + q + r = 4$, $6$ and $8$. The results, and their comparison to the lattice data, are displayed in Fig.~\ref{fig:chi468}. We find that the lattice data are well reproduced for $T \lesssim 160 \MeV$. However, it can be noted that while the agreement is reasonable, these data are typically affected by larger error bars than those of the second order fluctuations studied above, and the behavior turns out to be noisier.

\section{Semiclassical estimates of the baryonic susceptibilities}
\label{sec:semiclassical}

The computation of the $qD$ model spectrum in Sec.~\ref{subsec:baryon_spectrum} has been performed by considering numerical methods for the solution of the quantum mechanical two body system. We will study in this section the spectrum and the baryonic susceptibilities within a semiclassical expansion, and provide some analytical formulas valid at high enough hadron masses and/or temperatures. While Eq.~(\ref{eq:chi_HRGM}) is the basic formula that we will consider for the susceptibilities, we prefer to write it in the following form
\begin{equation}
\chi_{ab}(T) =  \frac{1}{2\pi^2} \sum_{\zeta=\pm} \int_0^\infty dM 
\rho^{\zeta}_{ab}(M) \left( \frac{M}{T} \right)^2 \sum_{k=1}^\infty \zeta^{k+1}K_2\left( \frac{kM}{T} \right) \,,
\label{eq:chi_HRGM2}
\end{equation}
where
\begin{equation}
\rho^\zeta_{ab}(M) =  \sum_{i } g_i \, q_i^a \, q_i^b \, \delta(M-M_i) \,,
\end{equation}
and the sum is over mesons or baryons for $\zeta = \pm 1$ respectively. Note that for the baryonic susceptibility $\chi_{BB}$, one has that $\rho_{BB}(M)$ is equal to the density of states $\rho(M) = dN(M)/dM$, as $B = \pm 1$ for (anti)baryons.

\subsection{Semiclassical expansion}
\label{subsec:semiclassical}

The spectrum of a quantum mechanical system, more specifically the density of states, can be computed in a derivative expansion, an approximation that is closely related to a semiclassical expansion (also called WKB method) in the high mass regime~\cite{Caro:1994ht}. Within the WKB method, the leading contribution to the cumulative number writes
\begin{equation}
N(M) = \Tr(\Theta(M-\hat{H})) \simeq  \int \frac{d^3x d^3p}{(2\pi)^3} \Theta(M - H)  + \cdots \,,  \label{eq:Ncum_WKB}
\end{equation}
where the (classical) Hamiltonian is given by~\footnote{The effect of the constant additive term~$\mu$ is just a shift $N(M) \to N(M-\mu)$, so that we disregard~$\mu$ in our explicit expressions in the following.} 
\begin{equation}
H = \sqrt{\vp^2 + m_q^2 } + \sqrt{\vp^2 + m_D^2} + \sigma r  -\frac{4\alpha_S}{3r} + \mu\,.
\end{equation}
The zeroth order term in Eq.~(\ref{eq:Ncum_WKB}) (which will be denoted by WKB$_0$) has been made explicit, while the dots stand for higher order contributions in the derivative expansion. The trace in the second equality of Eq.~(\ref{eq:Ncum_WKB}) is taken in the center of mass system subspace.

\subsection{Massless quarks and diquarks}
\label{subsec:massless}

Let us consider first the case of massless quarks and diquarks, and treat the Coulomb term perturbatively, i.e. $H = H_0 + H_1$ with 
\begin{equation}
H_0 = 2p + \sigma r \,, \qquad H_1 = -\frac{4\alpha_S}{3r} \,,
\end{equation}
and $p = |\vp|$. Using the Taylor expansion of the step function
\begin{equation}
\Theta(M - H) = \Theta(M - H_0) - \delta(M - H_0) H_1 + \frac{1}{2}\delta^\prime(M - H_0) H_1^2 + \cdots \,, \label{eq:Theta}
\end{equation}
and after performing the spatial and momentum integral of Eq.~(\ref{eq:Ncum_WKB}), one finds the following contribution of the cumulative number at leading order in the semiclassical expansion
\begin{equation}
N^{\WKB_0}(M)= \frac{M^6}{720 \pi \sigma^3} + \alpha_S \frac{M^4}{36 \pi
  \sigma^2} + \alpha_S^2 \frac{2M^2}{9\pi \sigma} 
+ \cdots \,.
\label{eq:N_semi}
\end{equation}
Plugging this result into Eq.~(\ref{eq:chi_HRGM2}) with $\zeta = -1$, one gets the baryonic susceptibility
\begin{equation}
\chi_{BB}^{\WKB_0}(T) =  \frac{127 \pi^5}{94500} \left( \frac{T^2}{\sigma} \right)^3 + \frac{31 \pi^3}{5670} \alpha_S \left( \frac{T^2}{\sigma} \right)^2 + \frac{7\pi}{405} \alpha_S^2 \frac{T^2}{\sigma} 
+ \cdots \,.
\end{equation}
A factor of 2 has to be included to account for the antibaryons.

\subsection{Finite mass effects}
\label{subsec:massive}

\begin{figure*}[t]
\centering
 \begin{tabular}{c@{\hspace{3.0em}}c}
 \includegraphics[width=0.43\textwidth]{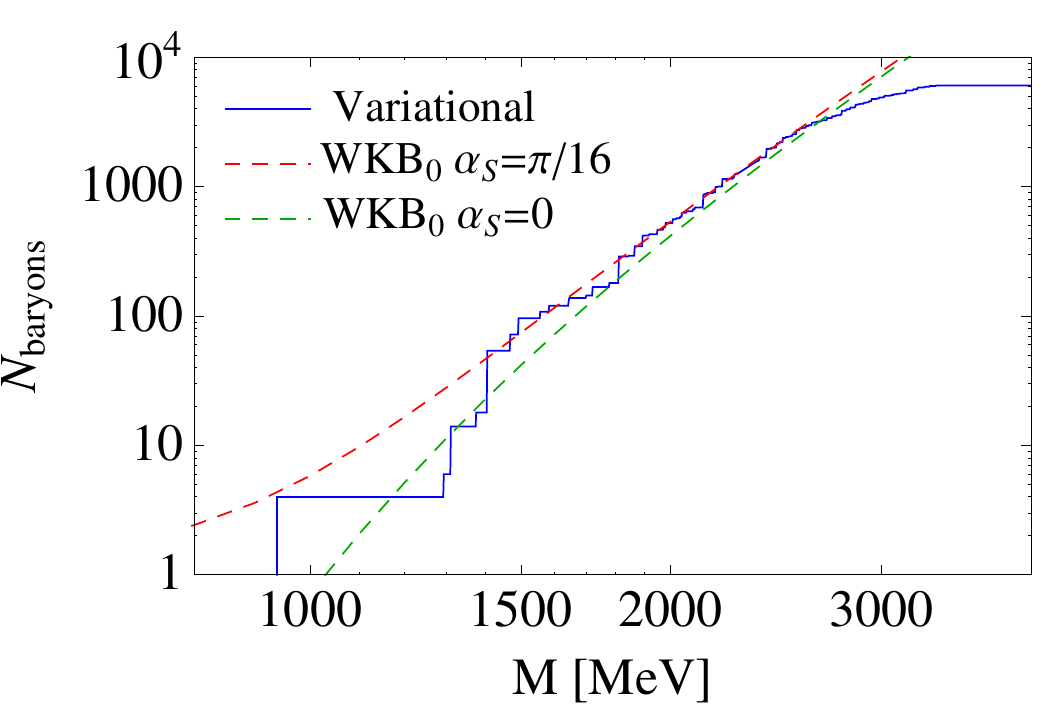} &
\includegraphics[width=0.43\textwidth]{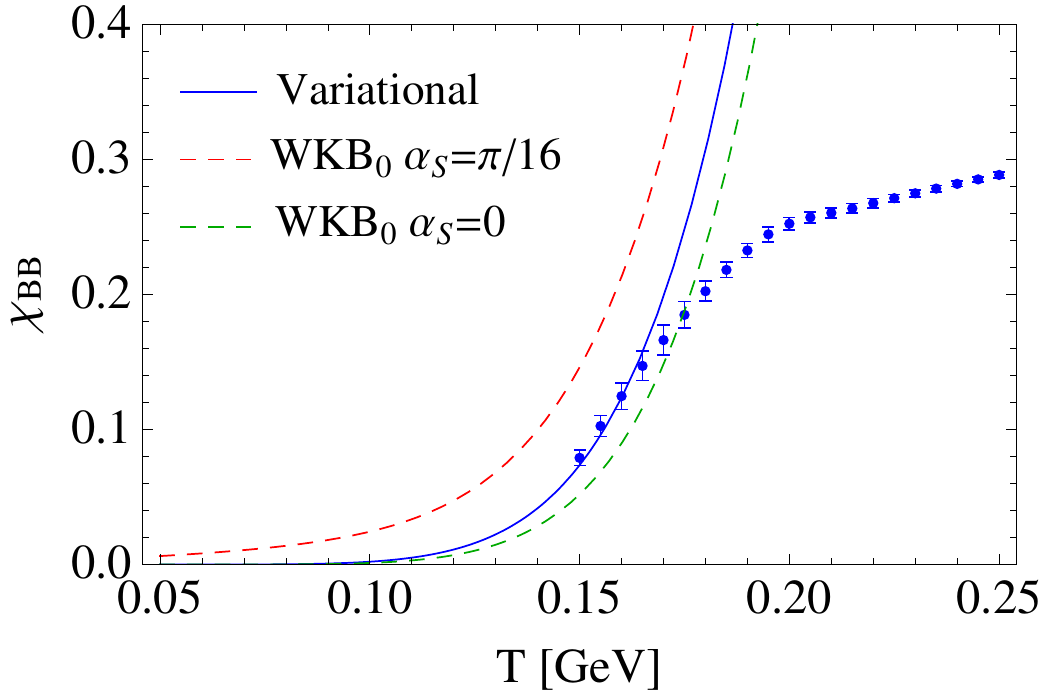} 
\end{tabular}
 \caption{\it Left panel: Cumulative number of the spectrum of baryons with the quark-diquark model. Right panel: Baryonic $\chi_{BB}$ susceptibility obtained with that spectrum. We display as dashed lines the result from the WKB approximation at leading order. The points in the right panel are the lattice data of Ref.~\cite{Bazavov:2012jq}. }
\label{fig:WKB}
\end{figure*}

If $\alpha_S$ is set to zero, the integral in Eq.~(\ref{eq:Ncum_WKB}) can be computed analytically in the massive case by considering the following change of variable in the momentum integral 
\begin{equation}
p \longrightarrow E := \sqrt{p^2 + m_q^2 } + \sqrt{p^2 + m_D^2} \,.
\end{equation}
The result for the cumulative number turns out to be rather lengthy, but we can provide a shorter expression corresponding to the asymptotic limit of  $M$ near and above the (classical) threshold $M_{\rm threshold}=m_q+m_D$, i.e.
\begin{equation}
N^{\WKB_0}(M) = \frac{64\sqrt{2}}{945 \pi \sigma^3} \left( \frac{m_q m_D}{m_q + m_D} \right)^{3/2} (M-m_q-m_D)^{9/2} +
\cdots \,.
\label{eq:N0largem}
\end{equation}
This region dominates the behavior of the susceptibility at small
temperatures, namely,
\begin{equation}
\chi_{BB}^{\WKB_0}(T) = \frac{(m_q m_D)^{3/2} T^3}{\pi^2 \sigma^3} e^{-(m_q + m_D)/T}  + \cdots \label{eq:chiBB_largem}
\end{equation}
for $T \ll m_q + m_D $.  This temperature regime is appropriate for the values
of $m_q$, $m_D$ and~$T$ considered in this work. We display in Fig.~\ref{fig:WKB} the result for the cumulative number and baryonic susceptibility computed with the WKB approximation presented above, and compared with the variational procedure of Sec.~\ref{subsec:baryon_spectrum} and the lattice data of Ref.~\cite{Bazavov:2012jq}. Notice that the WKB result correctly reproduces the baryon spectrum for masses $M \gtrsim 1500 \MeV$.

\section{Conclusions}
\label{sec:conclusions}

In the present work we have studied the baryon spectrum in the $(uds)$
flavor sector of QCD. We have argued that the asymptotic three-body
phase space for confined $qqq$ systems $\sim M^{12}$ is much larger
than the one actually observed in the PDG or determined in the RQM,
$\sim M^{6}$. This resembles a two body confined system despite the
fact that, by construction, baryons in the RQM are explicitly
described as $qqq$ bound states.  This strongly suggests a dynamics
for excited baryons dominated by quark-diquark degrees of freedom,
motivating the use of a quark-diquark model to compute the baryon
spectrum. In a simplified version of the model hyperfine splittings are
neglected and the pertinent quark-diquark potential is found to be
identical to the quark-antiquark using the appropriate Polyakov loop
correlators, and in strong agreement with lattice QCD
determinations. The obtained baryon spectrum and its related baryonic
susceptibilities in a thermal medium have been evaluated within the
HRG approach and successfully fitted to lattice QCD data in terms of
the constituent diquark mass and the current strange quark mass.  The
results fall in the bulk of previous intensive studies where a
detailed description of the spectrum was pursued. The extension of our
work to nonbaryonic susceptibilities would require a specific model
for mesons. These and other issues will be addressed in a forthcoming
publication~\cite{Megias:2019inprogress}.

\ack 
We would like to thank O.~Philipsen for enlightening discussions. This
work is supported by the Spanish MINECO and European FEDER funds
(Grants No. FIS2014-59386-P and FIS2017-85053-C2-1-P), by the FEDER/Junta de
Andaluc\'{\i}a-Consejer\'{\i}a de Econom\'{\i}a y Conocimiento
2014-2020 Operational Programme (Grant
No. A-FQM-178-UGR18), by Junta de Andaluc\'{\i}a (Grant No. FQM-225),
and by the Consejer\'{\i}a de Conocimiento, Investigaci\'on y
Universidad of the Junta de Andaluc\'{\i}a and European Regional
Development Fund (ERDF) (Grant No. SOMM17/6105/UGR). The research of
E.M. is also supported by the Ram\'on y Cajal Program of the Spanish
MINECO (Grant No. RYC-2016-20678).

\section*{References}

\end{document}